\documentstyle[11pt]{article}

\newlength{\abstractwidth}
\renewcommand{\thefootnote}{\fnsymbol{footnote}}
\renewcommand{\thanks}[1]{\footnote{#1}} 
\newcommand{\starttext}{
\setcounter{footnote}{0}
\renewcommand{\thefootnote}{\arabic{footnote}}}
\newcommand{\be}{\begin{equation}}
\newcommand{\br}{\begin{array}}
\newcommand{\er}{\end{array}}
\newcommand{\beq}{\begin{equation}}
\newcommand{\ee}{\end{equation}}
\newcommand{\eeq}{\end{equation}}
\newcommand{\ds}{\displaystyle}

\newcommand{\w}{\omega}
\newcommand{\<}{\langle}
\newcommand{\D}{\Delta}
\newcommand{\G}{\Gamma}

\renewcommand{\>}{\rangle}
\def\ba{\begin{eqnarray}}
\def\ea{\end{eqnarray}}
 
\begin{document}
\begin{titlepage}
\bigskip

\hskip 3.7in\vbox{\baselineskip12pt
\hbox{hep-th/9905186}}
\bigskip\bigskip\bigskip\bigskip

\centerline{\LARGE \bf
On the construction of local fields }
\bigskip
\centerline{\LARGE \bf
in the bulk of $AdS_5$ and other spaces.}

\bigskip\bigskip
\bigskip\bigskip

\centerline{Iosif Bena}
\bigskip

\centerline{\it Department of Physics}
\centerline{ \it University of California}
\centerline{ \it Santa Barbara, {\rm CA} 93117}
\centerline{iosif@physics.ucsb.edu}

\begin{abstract}

In the Poincar\'e patch of Minkovski $AdS_5$ we explicitly construct local bulk fields from the boundary operators, to leading order in $1/N$, following the ideas in \cite{bdhm,bklt}. We also construct the Green's function implicitly defined by this procedure. We generalize the construction of local fields for near horizon geometries of Dp branes. We try to expand the procedure to the interacting case, with partial success.

\baselineskip=16pt

\end{abstract}
\end{titlepage}
\starttext
\baselineskip=18pt
\setcounter{footnote}{0}


\section{Introduction}

In the context of the Maldacena conjecture \cite{maldacena,witten,gkp}, applied to the near horizon geometries of Dp-branes, we consider the possibility of constructing bulk fields from the boundary operators.
It is not obvious that local physics in a d+1 dimensional theory can be obtained from a d dimensional theory because of the different causal structures. More precisely, local fields in the higher dimensional theory should commute outside of the light cone, which contains one extra coordinate. Commutation of fields at spacelike separations along that direction is a nontrivial property.

In \cite{bdhm,bklt}, a method was presented for constructing fields in the bulk of $AdS_5$ from the operators of the boundary CFT at leading order in $1/N$. In \cite{bdhm}, the issue of locality was explored  thoroughly, but the arguments used made heavy use of the conformal structure of the boundary theory. Nevertheless, intuition tells us that it should be possible to construct local bulk fields in all the cases where the Maldacena conjecture applies - i.e. even in cases where the boundary theories are not conformal. Such cases are obtained by looking at Dp-branes with $p\neq 3$ and at their near horizon geometry.

Rather than using arguments having to do with conformal symmetry, we take the hard approach, explicit construction. This approach is less elegant, but easier to see through, and  has the advantage of easily extending to the nonconformal cases. We first do the construction for the $AdS_5$ - CFT case (in chapter 3) and then for one nonconformal case (in chapter 4).

We also construct the Green's function relating bulk fields with boundary operators, and explore its properties (in chapter 2). The Green's function is not very useful for checking locality for the bulk fields (unless you are a fan of hypergeometric functions), but is helpful at understanding the construction.
One interesting thing which we explore is the fact that the Green's function is not manifestly causal.

Finally, in chapter 5 we present a way to generalize this procedure to the interacting case.
Following the line of thought from \cite{bdhm} and our intuition, we believe that the theory in the bulk remains local in perturbation theory. Locality should be broken because of holography, but probably not perturbatively. We present a way to construct interacting fields and to check for their locality, at first order in $1/N$.  Nevertheless, checking for locality is easier said than done. We encounter computation difficulties too big for us. However, we will present the work done with the hope that some reader with more audacity and more technical inclinations might bring it to completion. 

\section{Construction of fields and of the Green's function on the Poincar\'e patch of $AdS_5$}

We briefly review the BDHM-BKLT \cite{bdhm,bklt} procedure. 
In a large $N$ conformal theory, we can obtain a set of chiral primary operators, with normalized orthogonal 2 point functions:
$$\<{O_i O_j}\> \sim \delta_{i\, j}, 
\eqno(2.1)$$
to leading order in $1/N$.  This looks like a free theory in which the $O_i$'s are the independent free fields. It is possible therefore to use their Fourier components, which behave as creation and annihilation operators to construct bulk free fields. 
A bulk field is constructed by multiplying the creation and annihilation operators by the appropriate normalized solutions of the bulk equation of motion.

In our case, the boundary is $R^3 \times R^1$, so the ``creation'' and ``annihilation'' operators are 
$$O_{{\bf k},\w} = {1 \over (2 \pi)^2}\int_{-\infty}^{+\infty } e^{ -i \w t - i \bf k x} O({\bf x},t) \, d^3{\bf x} \, dt,  
\eqno(2.2) $$ and its conjugate $O^{\dagger}_{{\bf k},\w}$. Here, $\w>0$.

The Poincar\'e patch of $AdS$ has the metric:
$$ds^2={1\over z_0^2}(-dt^2 + dz_0^2+d{\bf z}^2) 
\eqno(2.3)$$ 
The equation of motion for a field of mass $m$ is easy to obtain, and has 2 independent solutions: 
$$f^1_{\w,\bf k} = z_0^2 \, J_\nu (z_0 \sqrt{\w^2- k^2})  e^{ i \w t + i \bf k z} 
\eqno(2.4)$$ and
$$f^2_{\w,\bf k} = z_0^2 \, N_\nu (z_0 \sqrt{\w^2- k^2}) e^{ i \w t + i \bf k z}, 
\eqno(2.5)$$ 
where $\nu \equiv \sqrt{4+m^2}$, and $J$ and $N$ are Bessel functions of the first kind.
The boundary operator whose Fourier components are the creations and annihilation operators of a bulk field with mass $m$ has dimension $\D = 2+\sqrt{4+m^2}$.
The exact prescription \cite {bdhm,pol} relating bulk fields and boundary operators is :
$$ \lim_{z_0 \rightarrow 0}{\Phi_i(t,{\bf x},z_0)} = z_0^\Delta O_i (t,{\bf x}) .
\eqno(2.6)$$

Because of general properties of field theories, the operators  $O^{\dagger}_{{\bf k},\w}$ and $O_{{\bf k},\w}$ only exist for $\w > |{\bf k}|$. This is consistent with the fact that one can only obtain a solution of the equation of motion with the behavior in (2.6) for this range of $\w$. 
We can actually see that $f^1$ properly normalized should be used as the mode function. Thus, 
$$\Phi^i(t,{\bf z},z_0) = {z_0^2 \over (2 \pi)^2} \int _{\w > |k|} d^3{\bf k}\,  d \w \, 
e^{ i \w t + i \bf k z} \left({\sqrt{\w^2- k^2}  \over 2}\right)^{-\nu}
J_\nu (z_0 \sqrt{\w^2- k^2})  O^i_{{\bf k},\w} + h.c. 
\eqno(2.7)$$

This equation implicitly defines a Green's function, which due to the abundance of this type of things in the literature we will call "transfer function". Thus, 
$$\int_{boundary} \hspace{-.3in} G(z, b)O^i(b) = \Phi^i(z),
\eqno(2.8)$$ 
where $z$ and $b$ are the generic coordinates in the bulk, respectively the boundary. 
Hence
$$G({\bf z}, z_0,t,{\bf x},\tau)={2 z_0^2  \over  (2 \pi)^4}  {\rm Re}  \int _{\w > |k|} d^3{\bf k}\, d \w\, e^{ i \w (t-\tau) + i \bf k (z-x)}  \left({\sqrt{\w^2- k^2}  \over 2}\right)^{-\nu} J_\nu (z_0 \sqrt{\w^2- k^2}) .
\eqno(2.9) $$
We can see that because we eliminated the modes with $\w<|k|$, the transfer function will not be proportional to a $\delta$ function when $z_0 \rightarrow 0$.  Thus, it is different from the analytical continuation of the Green's function obtained in \cite{witten}. Also, as we will see, this is the reason for its lack of manifest causality.

Finding $G$ explicitly is not hard, just a bit messy. We can first perform the angular part of the integral, by choosing 
$\bf z-x \equiv \D x $ as our main axis. Thus, ${\bf k (z-x)} = k \D x \cos \theta $, where ${0 < \theta < \pi /2},\ \  -\infty <k< \infty$. The choice of range for $\theta$ and $k$ is a bit unusual, but makes the computation easier. Thus,
$$G = {2^{\nu+2} \pi z_0^2  \over  (2 \pi)^4 } \ {\rm Re} \int _{\w > |k|} d\w \int _{-\infty}^{\infty} {dk \, k \over i \D x}\, e^{ i \w \D t + i k \D x} {J_\nu (z_0 \sqrt{\w^2- k^2}) \over (\sqrt{\w^2- k^2})^\nu}. 
\eqno(2.10)$$
Substituting $k = A \sinh y, \w = A \cosh y$ we obtain after a few steps
$$G ={2^{\nu+2} \pi z_0^2  \over  (2 \pi)^4 } 
\int _0^{\infty } dA\ A^{2-\nu} J_\nu(A z_0) \ 
{\rm Re} \int _{-\infty}^{\infty} {dy \sinh y \over i \D x} 
e^{ i A(\D t \cosh y  + \D x \sinh y)}. 
\eqno(2.11)$$

The second integral can be evaluated differently when $\D x > \D t $ and when $\D t > \D x$. We will present
the first case only. The second one can be obtained through analytic continuation. Substituting 
$\D x = \sqrt {\D x^2 - \D t^2} \cosh s$, $\D t= \sqrt {\D x^2 - \D t^2} \sinh s$, and $l=y+s$ the second integral becomes after a few steps:
$$ \begin{array}{lcl}
 \ds {{\rm Re} \!\int _{-\infty}^{\infty}\! {dy \sinh y \over i \D x}
e^{ i A(\D t \cosh y  + \D x \sinh y )}} \hspace{-.1in} &=& \hspace{-.1in} \ds {{2  \over \sqrt {\D x^2 - \D t^2} } \int_0^\infty \!\!\! dl \sinh l \sin(A \sqrt {\D x^2 - \D t^2} \sinh l)}\\
\ds{\hspace{2in}} &=&\hspace{-.1in} \ds{ {2 \over \sqrt {\D x^2 - \D t^2}} K_1(A \sqrt {\D x^2 - \D t^2}  )}.
\end{array}
\eqno(2.12) $$ 
Thus
$$ G ={2^{\nu+3} \pi z_0^2  \over  (2 \pi)^4 \sqrt {\D x^2 - \D t^2} } 
\int _0^{\infty } dA\  A^{2-\nu} J_\nu(A z_0) K_1(A \sqrt {\D x^2 - \D t^2}). 
\eqno(2.13) $$
In the case $\D t > \D x$, $K_1$ simply becomes $H_1$. Here, $J,N,H$ and $K$ denote the appropriate Bessel functions. The integral is not hard to perform. Using \cite{grad} (6.576 3) we obtain
$$ G ={ z_0^{2+\nu}  \over  (\D x^2 - \D t^2)^2  \pi^3 \G(\nu +1)  } 
F \left({2,1;\nu+1, {- z_0^2 \over \D x^2 - \D t^2}}\right) . 
\eqno(2.14) $$

As it is well known, the hypergeometric function has poles at $0, \ 1$ and $\infty$, and can be analytically continued inside the light cone. 

We have obtained a transfer function which does not look causal.  This is because its limit as the bulk coordinate goes to the boundary is not proportional to a $\delta$ function.  Thus, we need to know the boundary field all over the Poincar\'e patch in order to obtain the bulk field at a point. This contradicts our intuition that an excitation of the boundary should propagate causally in the bulk. 
We can see that we can only obtain a causal looking transfer function by working in the 
universal cover of $AdS$. In that case, we can regard the Poincar\'e patch where the boundary operators are as one in the back light cone of the bulk point, and thus have manifest causality.

\section{Locality in the bulk - the conformal case}

In order to check the locality of the bulk fields defined in (2.7), we need to find the commutator of $\Phi$ with itself, and with
its time derivative. One attempt at checking locality would be to use (2.14), but the computation is hard, and obscures the physics going on. More intuitive is to go in small steps.
Thus, we would like to find the commutators $[ O_{{\bf k},\w}, O^\dagger_{{\bf k}',\w'}]$ and 
$[ O_{{\bf k},\w}, O_{{\bf k}',\w'}]$  in the boundary theory, and combine them with (2.7) in order to obtain the commutator of the field with itself and with its time derivative. To do this, we first have to find $[O({\bf x},t), O({\bf 0},0) ]$. This can be done in two ways: by analytically continuing from the Euclidean case, or by using K\"all\'en-Lehmann representation.

We know that for properly normalized $O$'s, in the Euclidean case, 
$$\<{O({\bf x},\tau), O({\bf 0},0) }\>= {1 \over (x^2+\tau^2)^\D}, 
\eqno(3.1)$$ 
where $\D$ is the dimension of the operator $O$. 
We can obtain the time ordered and the anti time ordered correlation functions by analytically continuing under or above the poles at 
$\tau = x$ and $\tau = -x$.
For $t>0$, $  [O({\bf x},t), O({\bf 0},0) ] = \<{T \,O({\bf x},t) O ({\bf 0},0) }\> - 
\<{A\, O({\bf x},t) O({\bf 0},0) }\> $. Thus
$$[O({\bf x},t), O({\bf 0},0) ] =  {-2 i \sin (\pi \D) \theta(t^2-x^2) {\rm sgn} (t) 
\over (t^2-x^2)^\D }.
\eqno(3.2)$$
The formula above may look a bit puzzling for integer dimension operators, nevertheless, it only tells us is that the commutator is nonzero only on the light cone.
 
We can also  note that for a free field of mass $m$,
$$ \<{\phi ({\bf x},\tau), \phi ({\bf 0},0)}\>_E  =  {K_1(m \sqrt{x^2+\tau^2})\over \sqrt{x^2+\tau^2}}, 
\eqno(3.3a)$$
$$[\phi ({\bf x},t),\phi ({\bf 0},0)]_M= {i \pi J_1(m \sqrt{t^2-x^2})\over \sqrt{t^2-x^2}}\theta (t^2-x^2) {\rm sgn}(t),
\eqno(3.3b)$$
where the labels $E$ and $M$ stand for Euclidean and Minkovskian. We can use the K\"all\'en-Lehmann representation:
$$ \<{O({\bf x},\tau), O({\bf 0},0) }\>_E = \ds{\int_0^\infty dm^2 \, F(m^2) \  
 \<{\phi({\bf x},\tau), \phi({\bf 0},0)}\>_E}, 
\eqno(3.4a)$$
$$[O({\bf x},t), O({\bf 0},0)]_M =  \ds{\int_0^\infty dm^2 \,F(m^2) \  [{\phi({\bf x},t), \phi({\bf 0},0)}]_M},
\eqno (3.4b)$$
 and extract $F(m^2)$ from (3.4a) and (3.1), plug it back in (3.4b) and reobtain (3.2). The integral in the K\"all\'en-Lehmann representation runs from $0$, because there is no mass scale in a conformal theory. 

Since we defined $O_{{\bf k},\w}$ to correspond to positive frequency, we can easily see that 
$[ O_{{\bf k},\w}, O_{{\bf k}',\w'} ] = 0 $, since their commutator contains a $\delta (\w+\w')$. The same is true for the $O^\dagger$'s.
By looking at the Poincar\'e invariance of the boundary theory, it is quite easy to see that the commutator $[ O_{{\bf k},\w}, O^{\dagger}_{{\bf k}',\w'}]$ will contain $\delta(\w-\w') \delta({\bf k - \bf k'})$. Since the theory is conformal, the product of $\delta$ functions must be multiplied by an appropriate power of
$\w^2 - k^2$. 
We can also compute  $[ O_{{\bf k},\w}, O^{\dagger}_{{\bf k}',\w'}]$ by using brute force. Combining (2.2) and (3.2) is straightforward, and after a few steps in which we use techniques similar to the ones on pag. 2, we obtain:
$$ [ O_{{\bf k},\w}, O^{\dagger}_{{\bf k}',\w'}] = (\w^2-k^2)^{\D-2} \delta(\w-\w') \delta({\bf k - \bf k'}) C, 
\eqno (3.5) $$
where 
$C$ is a numerical constant ($-\pi^3 \over 2^{2 \D -5} \G(\D) \G(\D-1)$ for the curious reader). We understand $\delta({\bf v})$ of a d-dimensional vector ${\bf v} $ to be the appropriate d-dimensional $\delta$ function.
 
Using (3.5) and (2.7) we can show that the bulk fields are local.
There are three things to check: 

1 - $[\Phi(t',{\bf z},z_0),\Phi (t',{\bf z'},z'_0)] = 0$, 

2 - $[\Phi(t',{\bf z},z_0), {\partial \over \partial t}\Phi (t',{\bf z'},z'_0)] \sim \delta(z_0-z'_0) \delta({\bf z -\bf z'})$

3 - $[\Phi(t,{\bf z},z_0), {\partial \over \partial t}\Phi (t',{\bf z'},z'_0)] = 0$ outside of the light cone.

In fact, condition 3 should follow from 2, because of the symmetry of the bulk, but it is nice to show how it works out. We can fix without loss of generality $t'$ and $\bf z'$ to $0$. Using (2.7) and (3.5) we have
$$[\Phi(0,{\bf z},z_0),\Phi (0,{\bf 0},z'_0)] \sim 
z_0^2  z_0'^2 \hspace{-.1in}\int _{\w > |k|} \hspace{-.2in} d^3{\bf k}\, d \w \, J_\nu (z_0 \sqrt{\w^2- k^2}) J_\nu (z'_0 \sqrt{\w^2- k^2})  (e^{ i \bf k z} - e^{-i \bf k z}) \sim 0. 
\eqno(3.6) $$
$$[\Phi(t,{\bf z},z_0),\dot\Phi (0,{\bf 0},z'_0)] \sim 
z_0^2 z_0'^2 \hspace{-.1in} \int _{\w > |k|} \hspace{-.2in}
d^3{\bf k}\, d\w \, \w J_\nu (z_0 \sqrt{\w^2- k^2}) J_\nu (z'_0 \sqrt{\w^2- k^2})  
\cos( {\w t+ \bf k z} ).
\eqno(3.7) $$
Using the same change of variable as at the evaluation of (2.9), we obtain
$$[\Phi,\dot\Phi] \sim {z_0^2 z_0'^2 \over z} \int _0^{\infty } dA \ A^3 
J_\nu(A z_0) J_\nu(A z_0')  {\rm Im} \int _{-\infty}^{\infty} dy \, \sinh y \cosh y\,
e^{ i A( z \sinh y  + t \cosh y)}. 
\eqno(3.8)$$
For $t=0$, it is easier to use (3.7), and replace $\w\, d\w$ by $A \ dA$. The integrals over $d^3{\bf k}$ and $dA$ separate, and the $d^3{\bf k}$ integral gives us $\delta({\bf z})$. We are left with 
$$ [\Phi,\dot\Phi] \sim {z_0^2 z_0'^2 \delta({\bf z})} \int _0^{\infty } dA \ A J_\nu(A z_0)  J_\nu(A z_0'). 
\eqno (3.9a)$$  
The integral is proportional to $\delta(z_0-z_0')$, by the orthogonality relation of Bessel functions. Thus, we have checked the second relation. We did not keep all the constants along the way for the sake of clarity. The exact answer is:
$$[\Phi,\dot\Phi]= z_0^3 \delta(z_0-z_0')  \delta({\bf z}){2 \pi^2 i\over \G(\D)\G(\D-1)} 
\eqno(3.9b)$$
 
In order to check the third relation we have to look at the range we are in. For $z>t$, we can make the same substitution as in (2.10): $z = \sqrt {z^2 - t^2} \cosh s$, and $t= \sqrt {z^2 - t^2} \sinh s$,
From (3.7) we obtain after making a change of variable: 
$$ \br{lcl}
\ds{{\rm Im} \int _{-\infty}^{\infty}  \hspace{-.1in} dy \sinh y \cosh y \, e^{ i A( \sinh y z + \cosh y t)}} \hspace{-.1in} &=& \hspace{-.1in} \ds {\cosh 2s  \int _{-\infty}^{\infty} \hspace{-.1in}  dy \sinh y \cosh y \sin (A \sqrt {z^2 - t^2} \sinh y)}\\ &=& \hspace{-.1in} \ds{{\cosh 2s \over z^2-t^2}\delta'(A)} .
\er
\eqno (3.10)$$
Clearly the commutator vanishes for this range,  because of the high power of $A$ in (3.8). 

For $z=t$, we will have an extra term of the form ${1\over A^2} {\delta'(\sqrt {z^2 - t^2})\over \sqrt {z^2 - t^2} } $ in the last line of (3.10) . The original term will vanish again because of the high power of $A$ in (3.8), and the extra term will also vanish for $z_0 \neq z_0'$, because of the orthogonality relation of Bessel functions.

For $t>z$, we substitute $t = \sqrt {t^2 - z^2} \cosh s$, and $z= \sqrt {t^2 - z^2} \sinh s$, and get
$$\br{l} 
\ds{{\rm Im} \int _{-\infty}^{\infty} dy \sinh y \cosh y\, e^{ i A( z \sinh y  + t \cosh y )}} \\ \hspace{1in}
\ds{= \sinh 2s  \int _{-\infty}^{\infty} dy \cosh 2y \sin (A \sqrt {t^2 - z^2} \sinh y)} \\ \hspace{1in}
\vspace{.2in}\ds{\sim J_2(A \sqrt {t^2 - z^2})} .
\er
\eqno (3.11)$$
Thus 
$$[\Phi,\dot\Phi] \sim {z_0^2 z_0'^2 \over z} \int _0^{\infty } dA\ A^3 J_\nu(A z_0) 
J_\nu(A z_0') J_2(A \sqrt {t^2 - z^2}).   
\eqno(3.12)$$
Using \cite{grad}(6.578-8), and remembering that $J_2 (x)=J_{-2}(x)$, we can see that the integral vanishes for
$|z_0-z_0'| >  \sqrt {t^2 - z^2}$, e.g. outside of the light cone, which is what was expected.

\section {The non conformal case}
We can look at the near horizon metric and dilaton for a collection of $N$ Dp branes, for a general p:
$$\br{rcl}
\ds{ds_p^2} &=& \ds{\alpha' \left[{
{U^{(7-p)/2}\over g_{YM} N^{1/2}}dx_\|^2 + {g_{YM} N^{1/2} \over U^{(7-p)/2} }(dU^2+U^2d\Omega^2_{8-p}) }\right]},\\
\ds{e^\Phi} &=& \ds{g_{YM}^2 \left({ U^{(7-p)/2}\over g_{YM} N^{1/2}
}\right)^{(p-3)/2}}.

\er
\eqno(4.1)$$
Here $g_{YM}$ is the coupling constant for the theory on the brane. 
Supergravity is a valid description for the bulk theory in the range where both the dilaton and the scalar curvature are small. In the case of D2 branes, which we will explore later, this is  $g_{YM}^2 N^{1/5} < U < g_{YM}^2 N$. A more general analysis of other cases can be found in  \cite{imsy}. We look at the limit of $g_{YM}$ and $N$ where this is all the bulk.

A massless scalar field $\phi$ minimally coupled to the Einstein metric, with $S_{8-p}$ angular momentum quantum number $l$, frequency $\w$ and  momentum $\bf k$ satisfies the classical equation of motion:
$$\phi''(U)+{8-p \over U} \phi'(U)+\left({-{m^2\over U^2}+ {g_{YM}^2 N (\w^2-k^2)\over U^{(7-p)}} }\right)\phi(U)=0,
\eqno(4.2)$$
where $m^2=l(l+7-p)$ is the mass of the Kaluza-Klein mode corresponding to $l$.

Changing the coordinate to $z_0 \equiv {1 \over U}$, the mode function vanishing at the boundary is obtained to be:
$$ f_{\w,\bf k}(z_0) = z_0^{7-p \over 2} J_\nu \left({{2 g_{YM}N^{1/2} \over 5-p} \sqrt{\w^2- k^2} z_0^{5-p \over 2}}\right)  e^{ i \w t + i \bf k z}, 
\eqno(4.3)$$
where $\nu \equiv {\sqrt {4 m^2 + (7-p)^2}\over 5-p} = {2l + 7-p \over 5-p}$.

We expect the construction to work in all the nonconformal cases, since by the conjecture \cite{maldacena} the Hilbert spaces are the same. Nevertheless, we will only illustrate the case of the D2 brane.
The machinery is very similar to the one we had to crank in the conformal case.
Thus, for a bulk scalar with angular momentum $l$, the Euclidean correlator of the corresponding boundary operators is \cite{aki}:
$$\<{O({\bf x},\tau), O({\bf 0},0) }\>= {B \over (x^2+\tau^2)^\D}, 
\eqno(4.4)$$
where $B \equiv {N^{(5+2l)/3}\over (g_{YM}^2)^{(l+1)/3}}$ and $\D \equiv (19+4l)/6$.
Quite clearly, 
$$ [ O_{{\bf k},\w}, O^{\dagger}_{{\bf k}',\w'}] = B' (\w^2-k^2)^{\D-3/2} \delta(\w-\w') \delta({\bf k - \bf k'}) , 
\eqno (4.4) $$ 
with $B' \sim B$ up to a factor of order unity, which the curious reader can compute in the same way as in the previous chapter.

The bulk field satisfying the correspondent of (2.6), with $\bf z$ understood as a 2 dimensional vector is:
$$\Phi^i(t,{\bf z},z_0) = {z_0^{5/2}  (2/3 g_{YM}N^{1/2})^{-\nu}\over (2 \pi)^{3/2}} \int _{\w > k} \hspace{-0.1in} d^2{\bf k} \, d \w \,
e^{ i \w t + i \bf k z} A^{-\nu}
J_\nu ({\scriptstyle{2 \over 3}} 
g_{YM}N^{1/2}
A z_0^{3 \over 2})  O^i_{{\bf k},\w} + h.c. ,
\eqno(4.5)$$
where $A \equiv \sqrt{\w^2- k^2} $. For simplifying the formulas, we will call $g \equiv {{2\over 3} g_{YM}N^{1/2}}$.
We have the same three things to check as in the conformal case. The first one follows exactly like in the conformal case. The second one involves calculating
$$[\Phi(t,{\bf z},z_0),\dot\Phi (0,{\bf 0},z'_0)] \sim 
(z_0 z'_0)^{5/2} g^{-2\nu} B \ {\rm Re} \int _{\w > |k|} \hspace{-0.1in} d^2 {\bf k}\, d\w \, \w 
J_\nu (g A z_0^{3 \over 2})
J_\nu (g A {z'}_0^{3 \over 2})
e^{ i \w t+i \bf k z }. 
\eqno(4.6) $$
We can easily do the angular integral, using $d^2 {\bf k} = k \, dk \, d\theta$, with $0<k<\infty , \ \ 0<\theta<\pi$. 
After that, we make the substitution $k=A\sinh y,\w=A\cosh y$ and obtain 
$$\br{l}
\ds{[\Phi,\dot\Phi] \sim (z_0 z'_0)^{5/2} g^{-2\nu} B  
\int _0^{\infty } dA \ A^3 J_\nu (g A {z'_0}^{3 \over 2})
J_\nu (g A z_0^{3 \over 2})} \\ \hspace{2in} \ds{\cdot
\int _0^{\infty}  \sinh y \cosh y J_0(A \sinh y z) \cos(A \cosh y t)}. 
\er
\eqno(4.7)$$
We will call the second integral $I$. Using \cite{grad} (6.738 2), we can see that $I$ vanishes for $z>t$, as expected. In the equal time case, $t=0$, $I \sim {\delta (A z)\over Az}$, so $[\Phi,\dot\Phi] \sim \delta({\bf z})\delta (z_0-z_0')$, where $\delta (z_0-z_0')$ came from the orthogonality relation of Bessel functions applied to (4.7).

For $z<t$, $I \sim {|t|\over (t^2-z^2)^{(3/4)}} A^{-1/2}
J_{3\over 2}(A\sqrt{t^2-x^2})$. Thus
$$[\Phi,\dot\Phi] \sim  {(z_0 z'_0)^{5/2}|t|\over (t^2-z^2)^{(3/4)}} g^{-2\nu} B  
\int _0^{\infty } dA \ A^{{3 \over 2}+1} J_\nu (g A {z'_0}^{3 \over 2})
J_\nu (g A z_0^{3 \over 2}) J_{3\over 2}(A\sqrt{t^2-x^2}).
\eqno(4.8)$$
This integral vanishes when 
$g{z'_0}^{3 \over 2}-g z_0^{3 \over 2} > \sqrt{t^2-x^2}$ \cite{grad}(6-578, 5 and 8). 
To see if this indeed means outside of the ``light cone'' of the bulk theory, we can look at the metric (4.1) and see that in differential from, a null trajectory:  $dt=dz_0\, z_0^{1\over 2}g_{YM} N^{1\over 2}$, is exactly the differential of $g{z'_0}^{3 \over 2}-g z_0^{3 \over 2} = t$.
Therefore, we constructed local fields in the bulk. It is possible to turn the argument around, and argue that the in order to have a bulk local theory, the boundary operators had to satisfy a relation similar to (4.6).

\section{Interactions}

We have constructed local fields only to leading order in $1/N$. It is true that locality should ultimately be broken in the bulk, because of holography, but we believe this to be a nonperturbative effect. Thus it should be possible in principle to construct bulk local fields perturbatively in $1/N$. So, we can ask the question:
Can we repeat the BDHM-BKLT construction for interacting cases ?

At next to leading order in $1/N$, the boundary theory no longer looks like a free theory. Rather, for $i \neq j$  (3.1) is replaced by:
$$O_i(x_i) O_j(x_j) =  {1\over N} \sum_k {\int {d^dx Q_{ijk}(x|x_1,x_2) O_k(x) }},
\eqno(5.1)$$
where $Q_{ijk}(x|x_1,x_2)$ can be obtained from the boundary 3-point function.

Also, the bulk theory will have extra terms in the Lagrangian, of the form
$$L_{int}=d_{ijk}\Phi_i \Phi_j \Phi_k . 
\eqno(5.2)$$

We expect (2.6) to be valid in the interacting case as well. The practical way to check locality to next to leading order in $1/N$ is to use the interacting bulk equations of motion for the boundary ``free'' operators, and compare the result with the one obtained using the free bulk equation of motion for the boundary ``interacting'' theory. If the $1/N$ contributions cancel outside of the light cone, the bulk fields are local.

Nevertheless, the computation is very hard. We tried to compute the commutator of bulk fields using the boundary ``interacting'' theory and the free equations of motion, but the technical difficulties were too big for us. As we said in the introduction, we cannot but hope that some reader with more audacity and more technical inclinations might bring it to completion.

\section{Conclusions}

There are two new results in this paper. The first one is the computation of the transfer function relating boundary operators with bulk fields, in the context of the  $AdS$-CFT correspondence. 

The second one is the expansion of the BDHM-BKLT procedure to nonconformal cases. We presented a method of constructing bulk fields in the near-horizon geometry of a collection of Dp branes, and verified locality explicitly for the D2 brane case.

{\bf Acknowledgments:} I would like to thank specially Prof.  Joseph Polchinski for his help and guidance through this project. I am grateful to Vijay Balasubramanian, Dorin Bena, Gary Horowitz, Daniel Z. Freedman and Albion Lawrence for very useful conversations, and to Aki Hashimoto and Nisan Itzhaki for allowing the use of their unpublished results. This work was supported in part by NSF grant PHY97-22022.

\end{document}